\documentclass[draft]{agujournal2019}
\usepackage{url} 
\usepackage{lineno}
\usepackage{multirow}
\usepackage{soul}
\draftfalse

\journalname{JGR: Space Physics}

\newcommand{\correction}[1]{#1}

\newcommand{\correctiontwo}[1]{#1}

\begin{document}

%
%


\title{Effect of a magnetospheric compression on Jovian radio emissions: in situ case study using Juno data}

%
%




\authors{C.~K.~Louis\affil{1,2},
C.~M.~Jackman\affil{1},
G.~Hospodarsky\affil{3},
A.~O'Kane~Hackett\affil{1,4},
E.~Devon-Hurley\affil{1,4},
P.~Zarka\affil{2,5},
W.~S.~Kurth\affil{3},
R.~W.~Ebert\affil{6,7},
D.~M.~Weigt\affil{1,8},
A.~R.~Fogg\affil{1},
J.~E.~Waters\affil{9},
S.~C.~McEntee\affil{1,4},
J.~E.~P.~Connerney\affil{10},
P.~Louarn\affil{11},
S.~Levin\affil{12},
S.~J.~Bolton\affil{6}}

\affiliation{1}{School of Cosmic Physics, DIAS Dunsink Observatory, Dublin Institute for Advanced Studies, Dublin 15, Ireland}
\affiliation{2}{Observatoire Radioastronomique de Nan\c cay, Observatoire de Paris, Universit\'e PSL, CNRS, University Orl\'eans, Nancay, France}
\affiliation{3}{Department of Physics and Astronomy, University of Iowa, Iowa City, Iowa, USA}
\affiliation{4}{School of Physics, Trinity College Dublin, Dublin, Ireland}
\affiliation{5}{LESIA, Observatoire de Paris, PSL Research University, CNRS, Sorbonne Universit\'e, UPMC University Paris 06, University Paris Diderot, Sorbonne Paris Cit\'e, Meudon, France}
\affiliation{6}{Southwest Research Institute, San Antonio, Texas, USA}
\affiliation{7}{Department of Physics and Astronomy, University of Texas at San Antonio, San Antonio, Texas, USA}
\affiliation{8}{Department of Computer Science, Aalto University, Aalto, Finland}
\affiliation{9}{Department of Physics and Astronomy, University of Southampton, Highfield Campus, Southampton, SO17 1BJ, UK}
\affiliation{10}{Space Research Corporation, Annapolis, MD}
\affiliation{11}{IRAP, Universit\'e de Toulouse, CNRS, CNES, UPS, Toulouse, France}
\affiliation{12}{Jet Propulsion Laboratory, Pasadena, California, USA}






\correspondingauthor{Corentin Kenelm Louis}{corentin.louis@dias.ie}




\begin{keypoints}
\item This paper provides a list of the Jovian magnetosphere boundary crossings by the Juno spacecraft from June 2016 to August 2022.
\item Jovian magnetospheric compressions lead to increased bKOM radio emissions (immediately) and DAM on the dusk sector (more than one rotation later).
\item nKOM radio emission appears later during relaxation phase of the compression.
\end{keypoints}

%
%

%
%


\begin{abstract}

During its polar orbits around Jupiter, Juno often crosses the boundaries of the Jovian magnetosphere (namely the magnetopause and bow shock).
From the boundary locations, the upstream solar wind dynamic pressure can be inferred, which in turn illustrates the state of compression or relaxation of the system. 
The aim of this study is to examine Jovian radio emissions during magnetospheric compressions, in order to determine the relationship between the solar wind and Jovian radio emissions. 
In this paper, we give a complete list of bow shock and magnetopause crossings (from June 2016 to August 2022), \correction{and the associated solar wind dynamic pressure and standoff distances inferred from \citeA{2002JGRA..107.1309J}}. We then select two \correction{sets of magnetopause crossings with moderate to strong compression of the magnetosphere for two case studies} of the response of the Jovian radio emissions. We \correction{confirm} that magnetospheric compressions lead to the activation of new radio sources. Newly-activated broadband kilometric emissions are observed almost simultaneously \correction{with} compression of the magnetosphere, with sources covering a large range of longitudes. Decametric emission sources are seen to be activated more than one rotation later only at specific longitudes and dusk local times.
Finally, the activation of narrowband kilometric radiation is not observed \correction{until} the magnetosphere is in its expansion phase.

\end{abstract}

\section*{Plain Language Summary}

%
%

%


%
%
%
%

\section{Introduction}
\label{sec:introduction}
\correction{Planetary} studies \correction{often} face the challenge of interpreting \textit{in situ} spacecraft \correction{observations} without the benefit of an upstream monitor revealing the prevailing conditions in the interplanetary medium. \correction{This is particularly true of the outer planets.} Radio emissions provide a direct probe of the site of particle acceleration and have potential to be used as a proxy for magnetospheric dynamics (see e.g., \citeA{2022FrASS...9.0279C} for Saturn; \citeA{2022JGRA..12730209F} for Earth). At Jupiter, the radio spectrum is composed of \correction{at least six} components, from low-frequency emissions, such as quasi-periodic (QP) bursts or trapped continuum radiation (from a few kHz to 10s of kHz), up to decametric (DAM) emissions ranging from a few MHz to 40 MHz \correction{\cite{1983phjm.book..285G, 1998JGR...10320159Z, 2021JGRA..12629435L}}.

In this study, we focus on 
three
types of radio emissions observable with Juno: 
narrowband kilometric (nKOM),
broadband kilometric (bKOM) and auroral DAM emissions (i.e., not induced by Galilean moons). 
The nKOM is attributed to a mode conversion mechanism producing emissions inside Io's torus at or near the local electron plasma frequency \cite{1982RvGSP..20..316B, 1983phjm.book..285G, 1988pre2.conf..245J, 1992pre3.conf..405R}.
The last two components (bKOM and DAM) are auroral emissions, 
produced by the cyclotron maser instability (CMI), near the local electron cyclotron frequency. The sources of these emissions are located \correction{on magnetic field lines of magnetic apex} \correctiontwo{(M-Shell)} between $10$ and $60$ (unitless distance of the magnetic field line at the magnetic equator normalized to Jovian radius 71492 km). These emissions are very anisotropic and beamed along the edges of \correction{a} hollow cone with an opening of $\sim75^\circ \pm 5^\circ$ to $\sim90^\circ$ with respect to the local magnetic field lines \cite{1994P&SS...42..919L, 1998JGR...10320159Z, 2006A&ARv..13..229T, 2017GeoRL..44.4439L, 2018GeoRL..45.9408L, 2019GeoRL..46..571I, 2019GeoRL..4611606L}.
 

The relation of the different components of Jupiter's radio emissions to both internal and external drivers is complex, as shown by several previous studies.
These studies show a relationship between some of the components and external (solar wind) or internal (rotation, magnetic reconfiguration) drivers. Recently, \citeA{2021JGRA..12629780Z} have re-analyzed data from Cassini’s flyby of Jupiter, and found that hectometric (HOM) and DAM emissions are dominantly rotation-modulated (i.e. emitted from lighthouse-like sources fixed in Jovian longitude), whereas bKOM is modulated more strongly by the solar wind than by the rotation (i.e. emitted from sources more active within a given Local Time sector). This last study \correction{extends} earlier results by \correction{\citeA{1983Natur.306..767Z,1987A&A...182..159G, 2008GeoRL..3517103I, 2011JGRA..11612233I}}.
\citeA{1998GeoRL..25.2905L}, using Galileo radio observations, have shown a sudden \correction{onset}, \correction{and increased intensity (up to $2\times10^{-7}$~V.m$^{-1}$.Hz$^{-1/2}$ at $5$~MHz)} of bKOM and DAM radio emissions, as well as the activation of new nKOM radio emissions, during periods of magnetospheric \correction{disturbance. They postulated large-scale energetic events as reconfigurations} of the magnetosphere and plasmasheet \correction{somewhat analogous to} terrestrial substorms. 
The results obtained by \citeA{2010A&A...519A..84E}, using Ulysses spacecraft data during the distant Jupiter encounter and Nan\c cay Decameter Array (NDA) data, show that \correction{non-Io DAM radio emissions} occur during intervals of enhanced solar wind dynamic pressure, but without any direct correlation between the emission duration or power versus the solar wind pressure \correction{or} the interplanetary shock Mach number. 
Using 50 days of observations from Cassini and Galileo, \citeA{2002Natur.415..985G} showed that HOM emissions were triggered by the arrival of interplanetary shocks at Jupiter.
\citeA{2012P&SS...70..114H, 2014P&SS...99..136H} have also shown that an increase of the solar wind pressure affects the non-Io-DAM radio emissions, using ground-based radio measurements \cite{2012P&SS...70..114H} and Cassini and Galileo radio and magnetic measurements \cite{2014P&SS...99..136H}. These two studies have compared the type of shocks with the region of source activation. There are two type of shocks \cite{2015JGR....2015JA021138K}: fast forward shocks (FFS) and  fast reverse shocks (FRS). These shocks are driven by solar coronal mass ejection\correction{s} (CME) or corotating interaction regions (CIR). The sudden explosion of \correction{a} CME, at a higher velocity than the ambient solar wind\correction{,} usually drives a FFS. As this fast CME expands into the solar system and overtakes the slower background solar wind, a compressed interaction region is usually formed, which is delimited by FFS \correction{o}n one side and FRS \correction{o}n the other side \cite{1976GeoRL...3..137S, 2006JGRA..111.7S01T}. A FFS is characterized by a sharp or discontinuous increase of the solar wind velocity, density, temperature and magnetic field amplitude. A FRS is characterized by an increase of the solar wind velocity, but a decrease of the solar wind temperature, density and magnetic field amplitude.
Both \citeA{2012P&SS...70..114H, 2014P&SS...99..136H} studies have shown that FFS trigger mostly dusk emissions, whereas FRS trigger both dawn and dusk emissions, with a time delay depending on the strength/direction of the interplanetary magnetic field (IMF). All the shock-triggered radio sources were found to sub-corotate (i.e. rotating slower than the rotation period of Jupiter) with a rate ranging from $50$\% to $80$\% depending on the intensity of the IMF. These rates could \correction{respectively} correspond to the extended and compressed states of the Jovian magnetosphere.

The above cited studies relied on sparse datasets (flybys or remote measurements) but the once-in-a-generation Juno dataset gives the opportunity for longer-term monitoring of the Jovian system and its radio response. In particular, the apojoves early in the mission, which took Juno out to radial distances of $\sim110$~R$_\mathrm{J}$ on the dawn side, place the spacecraft near the nominal magnetopause and bow shock locations, and afford the opportunity to sample snippets of in situ solar wind, as well as to determine the positions of the magnetospheric boundaries at various points in time. All the while, the Juno radio instrument is constantly monitoring the Jovian radio spectrum. In this study we utilise this unique dataset to explore the connection between the solar wind and Jupiter's radio emissions by presenting the first case study of its kind. 

Section \ref{sec:methodo} describes the datasets and processing methodology. Section \ref{sec:results} presents case \correction{studies} of the Jovian radio emission response to \correction{two moderate to strong} magnetospher\correction{ic} compressions inferred from multiple magnetopause crossings while Juno is on the outbound leg of its trajectory. \correction{Finally in Section \ref{sec:discussion}}, we summarise and discuss the results of this study and present the perspectives.

\section{Methodology}
\label{sec:methodo}

\begin{figure}  
    \centering
    \includegraphics[width=\textwidth]{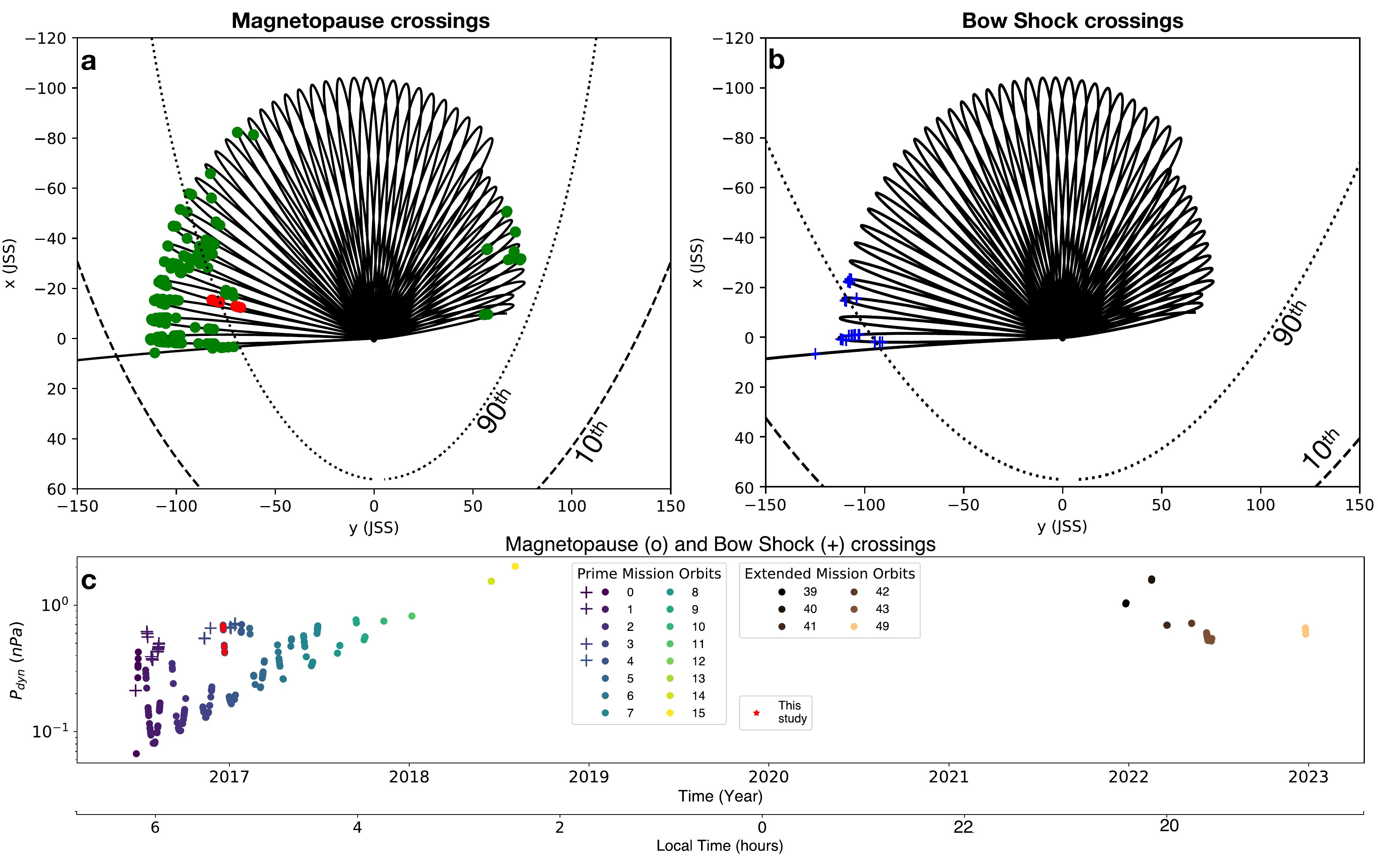}
    \caption{Projection \correction{of the Juno trajectory} into the equatorial plane, with the (a) magnetopause and (b) bow-shock crossings overplotted. The magnetopause crossings studied in this article are highlighted in red in panel (a). The coordinate system used here is the Jupiter-de-Spun-Sun (JSS). In this system, X points towards the Sun, Z is aligned with the Jovian spin axis, and Y closes the right-handed system (positive towards dusk).  \correction{In panel (a) the dashed line represents the $10^\mathrm{th}$ quantile position of the magnetopause ($0.03$~nPa), the dotted line its $90^\mathrm{th}$ quantile position ($0.518$~nPa). In panel (b) these same lines represent the $10^\mathrm{th}$ ($0.063$~nPa) and $90^\mathrm{th}$ ($0.579$~nPa) quantile positions of the bow shock \cite<values from>[]{2002JGRA..107.1309J}}. Panel (c) displays the solar wind dynamic pressure $P_\mathrm{dyn}$ values inferred from \citeA{2002JGRA..107.1309J}, for each crossing (``+": magnetopause; ``o": bow shock), as a function of time and Local Time (1200: direction of the Sun; 0000: opposition to the Sun). The colour code corresponds to the orbit number. The cases studied in this article are highlighted in red.}
    \label{fig:MS_boundary_crossing}
\end{figure}

Since July 2016, Juno has been in orbit around Jupiter, making a polar orbit every 53 days during its prime mission. Since the Ganymede flyby in June 2021, the orbits have been shortened to 43 days, before being reduced to 38 days in September 2022 with the Europa flyby. During its first 44 orbits, with an apojove of up to $\sim110$~R$_\mathrm{J}$, Juno crossed the boundaries of the magnetosphere several times \cite{2017GeoRL..44.4506H,2019JGRA..124.9106R,2022GeoRL..4999141M, 2020JGRE..12506366Collier}, as shown in Figure \ref{fig:MS_boundary_crossing} \correction{projected into} the equatorial plane. Figure \ref{fig:MS_boundary_crossing}a displays the magnetopause crossings while Figure \ref{fig:MS_boundary_crossing}b displays the bow shock crossings. In both of these panels are drawn the $10^{th}$ and $90^{th}$ quantile position of the magnetopause and bow shock, respectively, based on \correction{the} \citeA{2002JGRA..107.1309J} model. Note that this model was built on crossings from Ulysses, Voyager and Galileo, and thus may not be representative of all local times (especially the previously poorly explored dusk flank) or high-latitudes. The coordinate system used in this figure is the Juno-de-Spun-Sun (JSS), as this is the coordinate system used in the \citeA{2002JGRA..107.1309J} model. In this system, X points towards the Sun, Z is aligned with the Jovian spin axis, and Y closes the right-handed system (positive towards dusk). A 3D projection plot (in the Jupiter-Sun-Orbit (JSO) coordinate system) of the Jovian magnetosphere boundary crossings is shown in Figure S1 in Supporting Information (SI). In the JSO system, X is aligned with the Jupiter-Sun vector, Y indicates the Sun's motion in Jupiter frame, and Z closes the system.

\begin{figure}
    \centering
    \includegraphics[width=\textwidth]{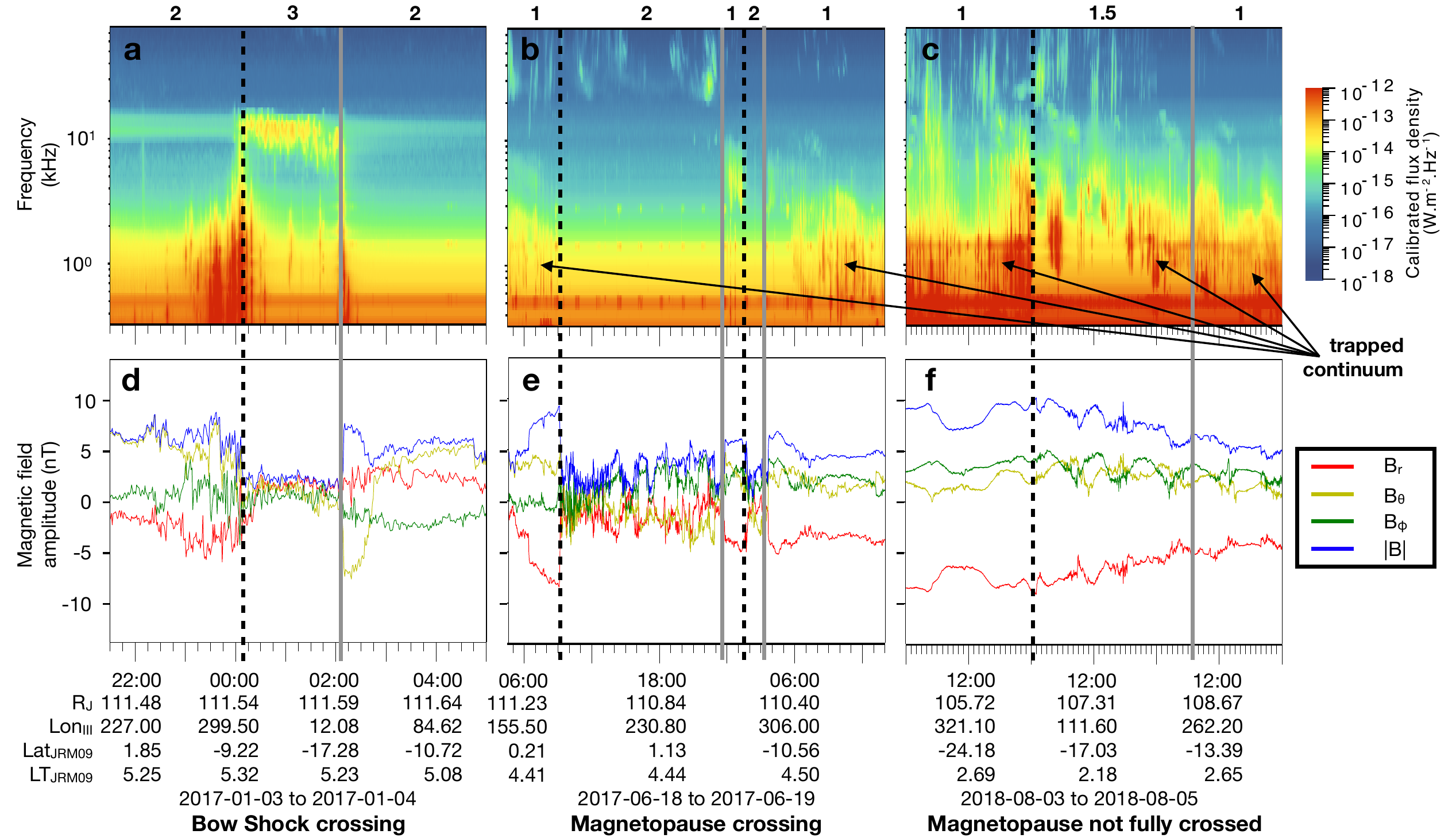}
    \caption{Examples of magnetospheric boundary crossings. Top panels (a-c) display Juno/Waves measurements \cite<using>[estimated flux density data set]{2021JGRA..12629435L, 2021Juno_Waves_Calibrated_data_collection}, while bottom panels (d-f) display Juno/MAG measurements in spherical JSO coordinates. Outbound crossings (boundary moving towards Jupiter) are highlighted by the black-dashed lines, while inbound crossings (boundary moving away from Jupiter) are highlighted by the grey-shaded lines. (left (a,d)) Bow shock crossings; (middle (b,e)) Magnetopause crossings; (right (c,f)) Example where the Juno spacecraft partially crossed the magnetopause without ever actually passing from the magnetosphere to the magnetosheath (i.e. moved around the border). The numbers above the Waves data indicate the region where Juno is located: (1): Magnetosphere, (2) Magnetosheath, (3) Solar Wind, (1.5): ``in'' the magnetopause boundary.}
    \label{fig:crossing_examples}
\end{figure}

In this study, the boundary crossings displayed Figure \ref{fig:MS_boundary_crossing} were determined using the radio measurements of the \correction{low frequency receiver} of the Juno/Waves instrument \cite{2017SSRv..213..347K}, and the magnetic field measurements of the Juno/MAG instrument \correction{using the Fluxgate Magnetometer measurements} \cite{2017SSRv..213...39C}, following the work done by \citeA{2017GeoRL..44.4506H}. Three examples are shown in Figure \ref{fig:crossing_examples}, with Juno/Waves data \cite<using>[estimated flux density data set]{2021JGRA..12629435L, 2021Juno_Waves_Calibrated_data_collection} displayed in the top panels, and Juno/\correctiontwo{MAG} data (in spherical JSO coordinates system) in the bottom panels. The ``out'' crossings (black dashed lines) correspond to a boundary moving towards Jupiter, e.g., Figure \ref{fig:crossing_examples}a,d, Juno crosses the bow shock going from the magnetosheath to the solar wind. The ``in'' crossings (grey shaded lines) define a boundary moving away from Jupiter, e.g., Juno crosses the bow shock, leaving the solar wind to enter the magnetosheath. 

The bow shock is a discontinuity formed when the supersonic solar wind is slowed to subsonic by interaction with the Jovian magnetic obstacle. A bow shock crossing is detected from the change in magnetic field amplitude and in the level of field fluctuations in the Juno/MAG data between the solar wind and the magnetosheath (Figure \ref{fig:crossing_examples}d). In the Juno/Waves measurements (Figure \ref{fig:crossing_examples}a) one can observe (i) an intense and broadband signal at the crossing and (ii) Langmuir waves when Juno is inside the solar wind, visible here at $\sim10$~kHz, which are produced by solar electrons reflected back into the solar wind from the shock boundary \cite{1971JGR....76.5162S, 1979JGR....84.1369F}.

The position of the magnetopause is determined by the balance between the solar wind dynamic pressure and the plasma pressure in the outer magnetosphere \cite{2004JGRA..109.9S12M}. A magnetopause crossing is detected by the appearance/disappearance in the Juno/Waves data \correction{(see Figure \ref{fig:crossing_examples}b)} of the trapped continuum radiation, usually observed between $0.5$~kHz and $2$~kHz. This signal is only seen when the observer is inside Jupiter's magnetosphere, \correction{in this example} before the black-dashed line at $\sim2017$-$06$-$18$T$09$:$00$, and after the grey-shaded line at $\sim2017$-$06$-$19$T$03$:$00$. This trapped continuum radiation propagates at a frequency lower than the plasma frequency inside the magnetosheath and therefore can not propagate into the magnetosheath (hence the name ``trapped"). Juno/MAG measurements of the magnetic field amplitude (Figure \ref{fig:crossing_examples}e) also show a change as Juno crosses the magnetopause, passing from the magnetosphere into the magnetosheath (see, e.g., black-dashed line at $\sim2017$-$06$-$18$T$09$:$00$), with \correction{a} decrease in magnetic field total amplitude $|B|$ and a much more disturbed signal than in the magnetosphere.

In some observations (see Figure \ref{fig:crossing_examples}c, between black-dashed and grey-shaded lines), low and high cut-off frequencies of the trapped continuum increase. Before $\sim2018$-$08$-$04$T$00$:$00$ (black-dashed line) and after $\sim2018$-$08$-$05$T$07$:$00$ (grey-shaded line), the trapped-continuum radiation is visible between $\sim0.3$~kHz and $\sim4$~kHz. In-between, the trapped-continuum radiation is no longer visible at low frequency, but is shifted to higher frequencies (between $\sim0.6$ and $\sim8$~kHz) and is very bursty. The high frequency part never completely disappears, and no drastic change in magnetic field components (Figure \ref{fig:crossing_examples}f) is observed, although they are more disturbed than in the magnetosphere, but less than in the magnetosheath. In the observation shown \correction{in} Figures \ref{fig:crossing_examples}c,f, Juno is on the outbound part of its trajectory and is therefore moving away from Jupiter. We interpret these observations as the movement of the magnetopause towards Juno at first (increase of low and high cut-off frequencies, see black-dashed line). Subsequently, the magnetopause stops moving towards Jupiter, and Juno never completely crosses the magnetopause to end up in the magnetosheath (between black-dashed and grey-shaded lines). 
Juno is however close enough to the magnetopause \correction{, or even in the boundary layer \cite{2011JGRA..116.4224W}}, to observe an \correction{increase of the low-frequency cutoff} of the trapped continuum by the increasing density when approaching the boundary.
Finally, the magnetopause is moving away from Jupiter (faster than Juno's velocity), and high and low cut-off frequencies decrease (Juno is again completely in the magnetosphere).

From the boundary positions, we can infer the solar wind dynamic pressure $P_\mathrm{dyn}$ using the \citeA{2002JGRA..107.1309J} model, by solving their second order polynomial equation \cite<equation 1 of>[]{2002JGRA..107.1309J}. \correction{From this, we can determine if the crossings of the magnetospheric boundaries are due to compressions of the magnetosphere, by comparing the inferred $P_\mathrm{dyn}$ values to either \citeA{2002JGRA..107.1309J} quantile values, or observed solar wind $P_\mathrm{dyn}$ distributions upstream of Jupiter \cite{2011SoPh..274..481J}.} \correction{One should note that the $P_\mathrm{dyn}$ value determined using Juno's position is not absolute, but a lower limit of the dynamic pressure. Although Juno is outbound, we cannot directly infer how far the magnetopause boundary is pushed back towards Jupiter.}

\correction{Figure \ref{fig:MS_boundary_crossing}c displays the inferred $P_\mathrm{dyn}$ for all crossings (``+": magnetopause; ``o": bow shock) as a function of time and Local Time. Note that there is a trend of increasing $P_\mathrm{dyn}$ values with time and decreasing Local Time. This is due to the procession of orbits, taking Juno more and more towards the night side of the magnetosphere (midnight Local Time), and thus deep into the magnetotail. This means that the magnetosphere has to be more compressed for Juno to cross the magnetospheric boundaries from this location. The bow shock is even further out again and thus Juno did not encounter the dawn side bow shock after the first few Juno orbits.}

In the absence of an upstream monitor, we can compare these inferred $P_\mathrm{dyn}$ values with those provided by solar wind propagation models \cite<e.g.,>[]{2005JGRA..11011208T}. For this, we must take into account any uncertainty on the propagation model values due to angle from opposition where predictions are most reliable. From this propagation model, we can also infer the type of shock (FFS or FRS) that compresses the magnetosphere \correction{as discussed in} Section \ref{sec:introduction}.

The full list of magnetopause and bow shock crossings (from 2016-06-24 to 2022-07-26, i.e. up to orbit 41) are available in Table S1 and S2 in Supplementary Information (SI), along with the position of Juno (in cartesian JSS --mandatory to use \citeA{2002JGRA..107.1309J} model-- and cartesian and spherical International Astronomical Union (IAU) System III  (SIII) coordinates system), the inferred solar wind dynamic pressure and the position of the magnetosphere standoff distances (bow shock and magnetopause) inferred from the \citeA{2002JGRA..107.1309J} model \cite{2022_boundary_crossings_lists}. \correction{Figure S2 displays statistical distributions based on the magnetosphere boundary crossings (Local Time, Solar Wind dynamic pressure, magnetopause and bow shock positions)}.

We next investigate the response of bKOM and DAM emissions \correction{to magnetospheric compression in a case study}. For that, we use \correction{the} \citeA{2021JGRA..12629435L} dataset \cite{2021Juno_Waves_Calibrated_data_collection} and catalogue of the radio emissions \cite{2021_Juno_Waves_catalog}. This catalogue contains the Jovian radio emissions identified in the Juno/Waves observations, \correction{only} from 2016-04-09 to 2019-06-24 (e.g. up to the 21$^\mathrm{st}$ apojove of Juno).  The radio components were visually identified according to their time-frequency morphology and then manually encircled by contours and labeled, using a dedicated program that records the coordinates of the contours and the label of each emission patch \cite{2022FrASS...901166L, 2022_SPACE_code_Louis}. While nKOM patches can be identified individually (fuzzy patches of emission elongated in time), the bKOM and DAM components have not been explicitly catalogued because they are the most frequent emissions in their respective frequency range. They can be selected and studied by excluding all other components and restricting to the adequate frequency range. For example, excluding nKOM in the range $20$-$140$~kHz allows \correction{one} to select the bKOM component only. In the [$3.5$-$40.5$]~MHz frequency range, only decametric emissions induced by the Galilean moons Io, Europa and Ganymede have been labelled 
(based on \citeA{2019A&A...627A..30L} simulations of those radio emissions, see \citeA{2020ExPRES_simulations_data_collection} for more details). Therefore, by excluding them, only auroral DAM emissions remain in this range. \correction{Given that HOM emissions can extend up to a few MHz, the highest part of the hectometric emission could be present in this range, but would only represent a minority of the emissions observed}.

\correction{For the case studies described in Section \ref{sec:results}, we decided to select the magnetopause crossings that took place between 2016-12-19 and 2016-12-23, highlighted in red in Figure \ref{fig:MS_boundary_crossing}. This choice is based on} \correctiontwo{three} \correction{factors: (i) in 2016-2017, the Jovian Auroral Distributions Experiment \cite<JADE,>[]{2017SSRv..213..547M} was not activated during excursions into the solar wind, excluding \textit{in situ} plasma information, and thus a direct measurement of $P_\mathrm{dyn}$. Therefore, we decided to choose among one of the (more numerous) magnetopause crossing cases; (ii) the case chosen had to be within the time interval covered by the catalogue of \citeA<>[i.e. between 2016-04-09 and 2019-06-24]{2021JGRA..12629435L}; (iii) in order to avoid any bias related to an extremely exceptional case, we did not select the case with the highest $P_\mathrm{dyn}$ value (second half of 2018, orbit 15).}

\correction{The time interval chosen presents} \correctiontwo{two main} \correction{advantages. (i) There are two sets of crossings in a row. The $P_\mathrm{dyn}$ value determined for the first crossing (2016-12-19T01:50) is $0.70$~nPa. The dynamic pressure associated with the second set of crossings (2016-12-21T08:48) is $0.48$~nPa. The distribution of $P_\mathrm{dyn}$ at Jupiter published by \citeA<>[see their Figure 4b]{2011SoPh..274..481J} reveals a peak at $0.05$~nPa and a maximum slightly above $1$~nPa. The $0.48$ and $0.70$ values therefore lie towards the tail of this distribution.
Moreover, these inferred values are close to the $90^\mathrm{th}$ quantile value ($0.518$~nPa) of the magnetopause position given by \citeA{2002JGRA..107.1309J}. Therefore, these two sets of magnetopause crossings correspond to a strong and a moderate compression.} \correctiontwo{(ii)} \correction{Based on Figure \ref{fig:MS_boundary_crossing}c (red points) the $P_\mathrm{dyn}$ values associated with these magnetopause crossings are well above the ``trend", and therefore correspond to the strongest compressions during orbit $4$. Recall that this ``trend" is due to the procession of Juno's orbit, taking the spacecraft deep into the magnetotail, implying that the magnetosphere needs to be more compressed for Juno to cross the magnetospheric boundaries.}

\section{Jovian auroral radio emission response to compressions of the magnetosphere}
\label{sec:results}

\subsection{Determination of the compression}
\begin{figure}
    \centering
    \includegraphics[width=\textwidth]{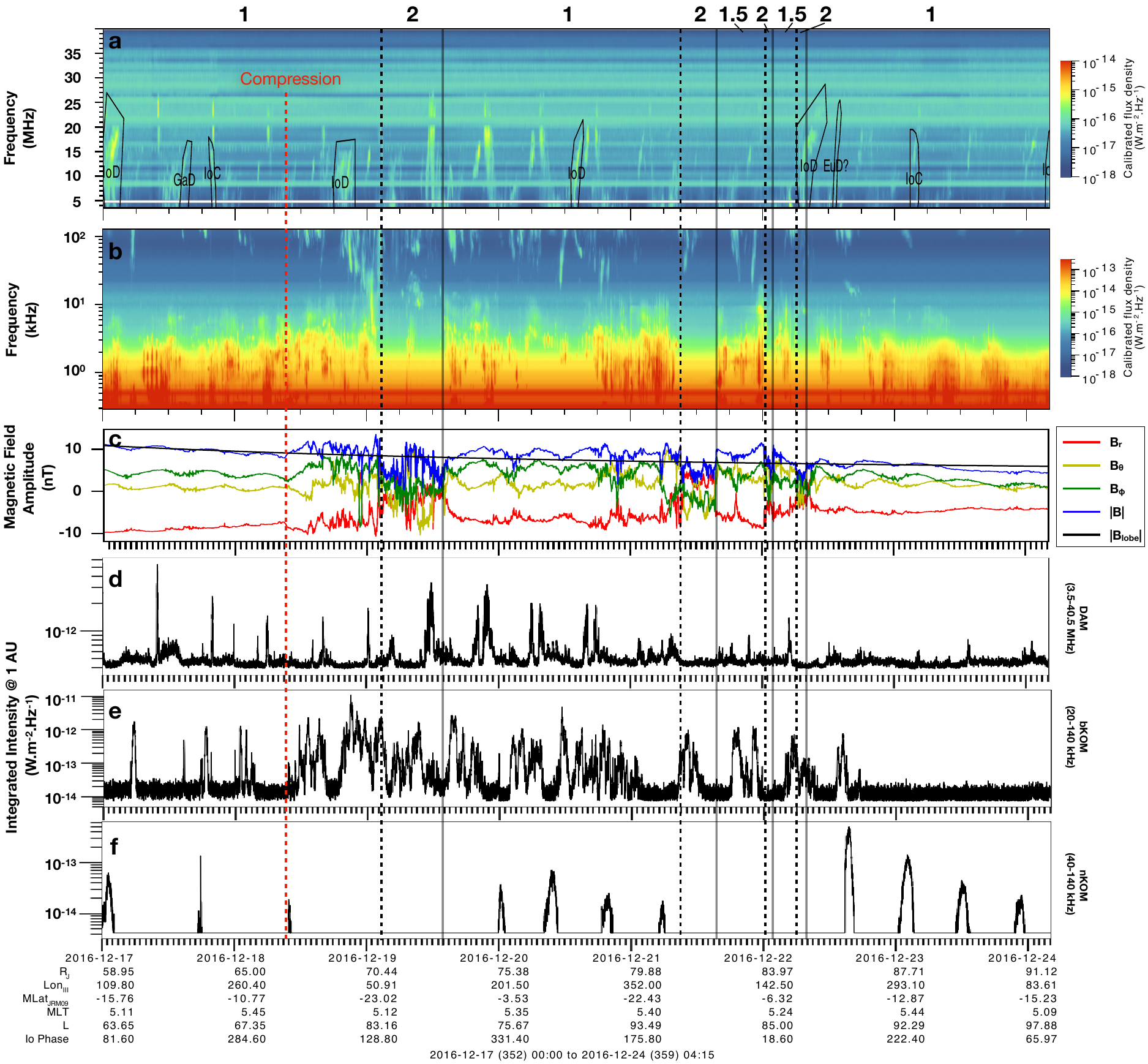}
    \caption{(a-c) Juno Waves and MAG measurements during a series of magnetopause crossings. Panels (a) and (b) show Juno Waves frequency-time spectrograms covering two different frequency ranges (from $3.5$ to $40.5$~MHz and between $3$~kHz and $140$~kHz, respectively), with the black polygons in the top panel denoting the radio emissions induced by the interaction between Jupiter and its moons \cite<e.g. Io, Europa or Ganymede, based on >[]{2021_Juno_Waves_catalog}. Panel (c) shows the three components of magnetic field (in JSO spherical coordinates system, red, yellow and green lines) and total field strength (blue). The black line displays the \citeA{2002JGRA..107.1196K, 2004jpsm.book..593K} magnetic field variation fit. Panels (d-f) display time series of integrated flux density (normalized at 1 Astronomical Unit (AU), $15$~sec time resolution) for (d) the auroral decametric (DAM, in the $3.5$-$40.5$~MHz range) not induced by the interaction between Jupiter and its moons (i.e. all the non-labelled emissions), (e) broadband kilometric (bKOM, in the $20$-$140$~kHz range)  and (f) narrowband kilometric (nKOM, in the $40$-$140$~kHz range) radio emissions. The black-dashed lines represent the outbound magnetopause crossings (from the magnetosphere to the magnetosheath) while the grey-shaded lines represent the inbound magnetopause crossings (from the magnetosheath to the magnetosphere). The red-dashed line represents the time when Juno starts to measure magnetic fluctuations and $|B| > |B_\mathrm{lobe}|$ (panel c), and an increase in the low and high cut-off frequencies of the trapped-continuum radiation (panel b). The numbers above the Waves data indicate the region where Juno is located: (1): Magnetosphere, (2) Magnetosheath, (1.5): ``in" the magnetopause boundary.}
    \label{fig:MP_crossing}
\end{figure}

Figure \ref{fig:MP_crossing} displays Juno measurements during magnetopause crossings for a 7-day interval from $2016$-$12$-$17$T$00$:$00$ to $2016$-$12$-$24$T$04$:$15$.
Black-dashed lines show when Juno crossed the magnetopause from the magnetosphere to the magnetosheath (outbound crossings), while grey-shaded lines show inbound crossings.
\correctiontwo{Figures} \ref{fig:MP_crossing}a,b display Juno/Waves measurements for two different frequency ranges: (a) [3--40.5]~MHz and (b) [0.3--140.0]~kHz. Figure \ref{fig:MP_crossing}c displays Juno/MAG measurements: total amplitude $|B|$, and ($r$,$\theta$,$\phi$) components in JSO spherical coordinates system. The black line displays the \citeA{2002JGRA..107.1196K, 2004jpsm.book..593K} magnetic field variation fit in the lobes (beyond $r=30$ Jovian radii, the lobe magnetic field falls off as $B_\mathrm{lobe} \mathrm{(nT)} = (2.94 \pm 0.07) \times 10^3$~$r^{-1.37\pm0.01}$). Therefore, for an observer inside the magnetosphere, and if the magnetosphere is in a steady-state, $|B|$ should follow $|B_\mathrm{lobe}|$.
\correctiontwo{Figures} \ref{fig:MP_crossing}d,e,f display integrated time series of the radio signal measured by Juno/Waves for three different radio components: (d) auroral DAM (i.e. excluding the \correction{satellite-related} DAM emissions), (e) bKOM, and (f) nKOM.

As described in Section \ref{sec:methodo} (see \correctiontwo{Figures} \ref{fig:crossing_examples}b,e), the magnetopause crossings are clearly seen in Figure \ref{fig:MP_crossing}b from the disappearing of the trapped continuum radiation and in Figure \ref{fig:MP_crossing}c from the change in the magnetic field components and total amplitude (see the black-dashed and grey-shaded lines). 
Looking in more detail at Juno/MAG measurements (Figure \ref{fig:MP_crossing}c), one can notice at $\sim 2016$-$12$-$18$T$09$:$00$ (indicated by the red dotted line), i.e., $\sim 18$~h before the crossing of the magnetopause, an increase of the $|B|$ (blue curve) and $B_\phi$ (green curve) components while the $B_\mathrm{r}$ (red curve) and $B_\theta$ (yellow curve) components decrease. This is followed by turbulence observed in all magnetic field components, but without the sharp decrease in $|B|$ characteristic of magnetic measurements in the magnetosheath. We also see, approximately at the same time, that the cut-off frequencies of the trapped continuum are increasing (Figure \ref{fig:MP_crossing}b): the trapped continuum is observable in the [$\sim0.4$-$3$]~kHz frequency range before the red-dashed line, and in the  [$\sim0.8$-$5$]~kHz frequency range between the red-dashed and black-dashed lines. \correction{This change in the cut-off frequencies is due to the inward motion of the magnetopause during the compression. Because of this, the local density along Juno's path is increasing, and therefore the low-frequency part of the trapped continuum cannot propagate}, resulting in an increase in the cut-off frequencies of the trapped continuum. All these characteristics are the signature of the inward motion of the magnetopause boundary towards the spacecraft (see \correctiontwo{Figures} \ref{fig:crossing_examples}c,f).

Furthermore, comparing the total amplitude of the magnetic field  $|B|$ (blue curve) to \citeA{2002JGRA..107.1196K, 2004jpsm.book..593K}  magnetic field variation fit $|B_\mathrm{lobe}|$, one can see that before $\sim 2016$-$12$-$18$T$09$:$00$ (red dotted line), $|B|$ and $|B_\mathrm{lobe}|$ follow the same trend. However, between  $\sim 2016$-$12$-$18$T$09$:$00$ and the crossing of the magnetopause (first black dashed-line), $|B|$ is above $|B_\mathrm{lobe}|$, which is a clear sign that the magnetosphere is being compressed \cite<see e.g.,>[]{2010JGRA..11510240J}.

All these elements lead us to interpret this as representative of the beginning of the impact of a stronger solar wind on the magnetosphere, and thus the beginning of compression. On the other hand, after Juno crosses the magnetopause for the second time (back into the magnetosphere, grey-shaded line) on $\sim 2016$-$12$-$19$T$14$:$12$ and until the next outward crossing of the magnetopause ($\sim 2016$-$12$-$21$T$08$:$48$), we observe the same features: a variable low and high cut-off frequencies of the trapped continuum, small perturbations in the magnetic field components, and $|B| > |B_\mathrm{lobe}|$. We interpret this as the relaxation phase of the magnetosphere, but not to a fully extended state. From the observations, \correction{we can deduce that Juno remains very close to the} magnetopause (same characteristics as in \correctiontwo{Figures} \ref{fig:crossing_examples}c,f), before the second compression takes place and the spacecraft is again in the magnetosheath.

\correction{By comparing the time spent by} Juno inside the magnetosheath \correction{during the two compression events}, we can infer whether one of the compressions was stronger than the other, i.e., lasted longer or the magnetopause was pushed further inwards. During the first pass from the magnetosphere to the magnetosheath, Juno stayed in it for $\sim 12$~h~$20$~min, whereas during the second pass, Juno stayed inside the magnetosheath less than $7$~h, before going back into the magnetosphere very quickly twice for a few minutes. Therefore, we can deduce that the first compression either lasted longer or the magnetopause was pushed further inwards. In any case, we can infer that the magnetosphere was probably more disturbed by the first compression.

The \citeA{2005JGRA..11011208T} solar wind propagation model is more reliable when Earth and Jupiter are in conjunction as seen from the Sun (Jupiter-Sun-Earth angle equal to $0^\circ$). During the time range displayed in Figure \ref{fig:MP_crossing}, the Jupiter-Sun-Earth angle is $-110^\circ$ (in average). Therefore, the error in timing on \citeA{2005JGRA..11011208T} solar wind propagation model can be as large as 2 days or more, the time interval between the shocks can also be shifted, and $P_\mathrm{dyn}$ can be misjudged. Therefore, the outputs from the \citeA{2005JGRA..11011208T} model should be used here only as a guide. For that reason, they are only displayed in the SI (Figure S3-S4), for information. According to \citeA{2005JGRA..11011208T} model, two shocks arrive at Jupiter successively in a time interval of two and a half days. The model predicts the arrival of the first compression at the beginning of day 2016-12-16, i.e. two days before the first compression observed by Juno. By shifting the model outputs by two days (see Figure S4), we obtain a good match between the arrival of the two shocks at Jupiter and the compressions observed by Juno.
These two shocks have very different characteristics (see Figure S3): (i) the first one shows an increase in the solar wind speed and a sharp decrease in the solar wind density and temperature, while (ii) the second shock shows an increase in the solar wind speed, density and temperature. Thus, if we take the outputs of \citeA{2005JGRA..11011208T} model as reliable, the first shock would be a Fast Reverse Shock (FRS) while the second would be a Fast Forward Shock (FFS). 


\subsection{Response of the auroral radio emission to the first compression}
Having determined the start time of the compression and the associated dynamic pressure, let us now study the response of the radio emissions to the first compression.

\subsubsection{Broadband kilometric (bKOM) emission}
The bKOM emissions (Figure \ref{fig:MP_crossing}e) are the first to show a strong variation. Before the onset of the compression, we can see some peaks in the integrated intensity, but restricted to a narrow frequency range (few 10s of kHz, see Figure \ref{fig:MP_crossing}b). Immediately after (dashed-red line at $\sim2016$-$12$-$18$T$09$:$00$), we observe emissions almost continuously, with an increase in the integrated intensity. This increase can be explained by both the observations of bKOM emissions over a much wider frequency range, i.e. from 20~kHz to 140~kHz (see Figure \ref{fig:MP_crossing}b), and by the increase intensity of the emission.
Very low frequency extensions of the emission, \correction{i.e., emissions extended} down to $20$~kHz, are only visible over $\sim1$~h~$15$~min, thus only for specific sources.
The bKOM emissions seen at \correction{almost} every longitude are then observed until $\sim2016$-$12$-$21$T$02$:$00$, thus over more than $60$~hours. \correction{The observation of emissions on an almost continuous basis tells us  that sources have been activated at almost all longitudes. It should be noted that no bKOM emissions seem to be observed between 2016-12-18T17:00 and 19:00. A sector of longitude therefore seems to have no associated bKOM emissions, at least during the first rotation. This could be due to various reasons, such as emissions that are too weak to be detected, geometric effects preventing the emission from being beamed towards the observer, or a sector that is completely non-activated.}

\subsubsection{Decametric (DAM) emission}
After compression, an increase in the integrated intensity of the DAM radio emissions is also observed. However, unlike the bKOM emissions, this is not observed simultaneously with the onset of the magnetic disturbances, nor is it continuous over time.
DAM emissions visible before the compression \correction{\cite<non-labelled vertex early arc up to 15 MHz, see Figure \ref{fig:MP_crossing}a, statistically reported by>[]{2017GeoRL..44.4584I}} are still visible during the compression with the same rotation period, however their intensity has increased compared to before the compression. \correction{Therefore, the appearance of these emissions is probably modulated by rotation and independent of any compression. However, compression seems to have an impact on their intensity.} New emissions, more intense and extending up to 25-30 MHz, appear at $\sim 2016$-$12$-$19$T$12$:$00$, i.e. $\sim 28$~hours after the compression, and last for $\sim 30$ hours. Their rotation period is longer than the previously visible DAM emissions, \correction{visible with} the double peak in the integrated time series Figure \ref{fig:MP_crossing}d, which means that the sources are \correction{sub-corotating} (see below).

\begin{figure}
    \centering
    \includegraphics[width=1.0\textwidth]{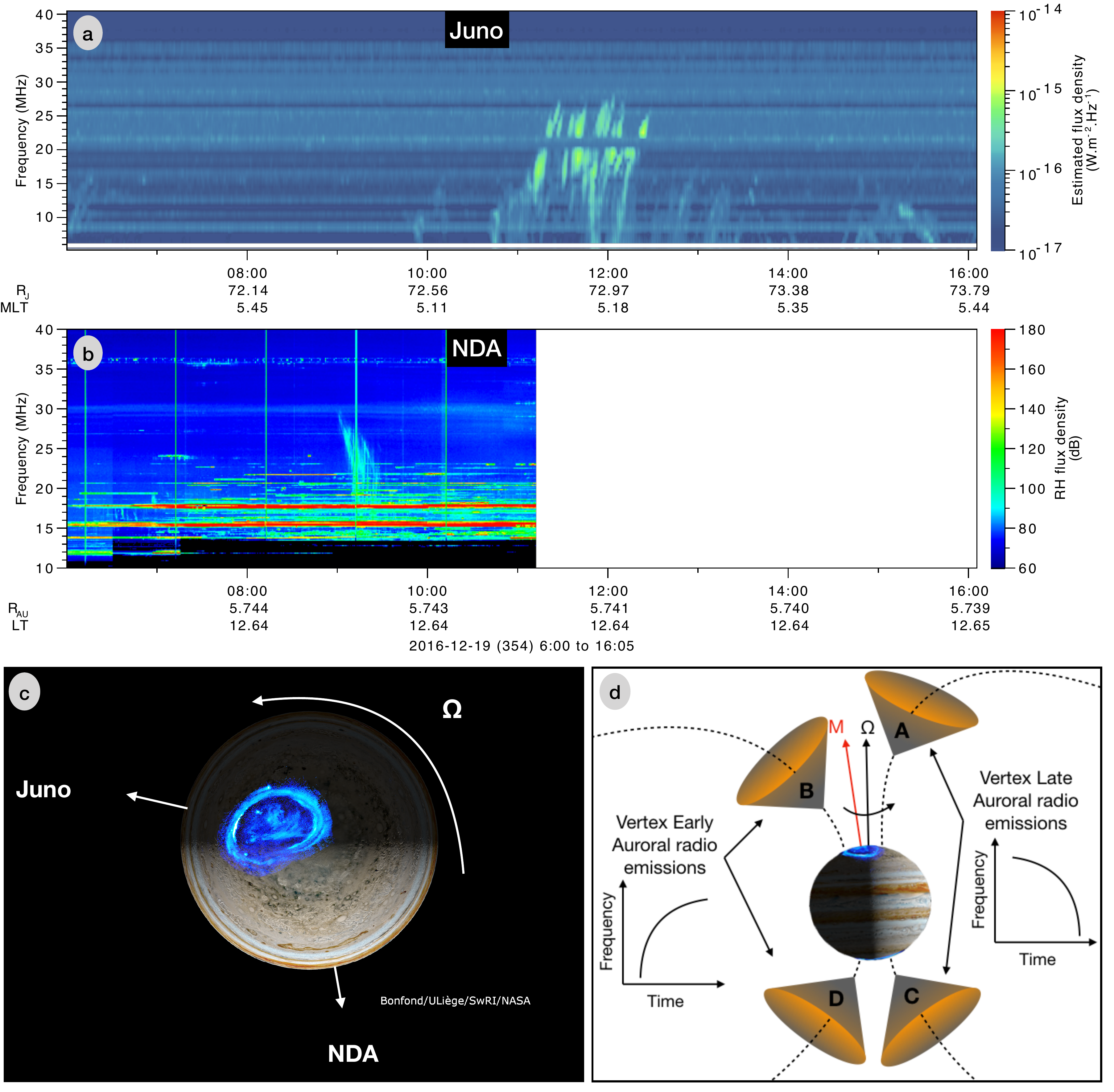}
    \caption{\correction{(\correctiontwo{a}) Juno Waves and (\correctiontwo{b}) Nan\c cay Decameter Array (NDA) routine receiver observations. Decametric radio emissions are clearly visible (\correctiontwo{a}) between 11:00 and 12:30 (Spacecraft Event Time) and (\correctiontwo{b}) between 09:00 and 09:30 (UT time). The light travel time between Juno and Earth is $\sim$ $47$~minutes. The data gap after 11:10 is due to the fact that Jupiter is no longer visible in the sky from the NDA observatory. (\correctiontwo{c}) Observers' configuration. (\correctiontwo{d}) Cartoon of the geometry and nomenclature of the auroral radio emissions and corresponding arc-shape in the (time, frequency) plane.} \correctiontwo{ If the source is located to the West of Jupiter for the observer (sources ``B" or ``D"), the emission will have a vertex early arc shape. If on the contrary the source is located to the East of Jupiter for the observer (sources ``A" or ``C"), the emission will have the shape of a vertex late arc.} The arcs observed in panels \correctiontwo{(a)} and \correctiontwo{(b)} originate from the same source. NDA sees the emission cone exiting its field of view (vertex late arc) while Juno sees the emission cone entering its field of view (hence vertex early arc).}
    \label{fig:NDA_observation}
\end{figure}

Since the CMI emissions are not isotropically emitted, but only emitted at the edge of a hollow cone, with an angle of $\sim 75^\circ$ to $\sim90^\circ$ with respect to the local magnetic field line (see Section \ref{sec:introduction}), geometry effects are important, and emission is mostly seen by an observer when the sources are at a longitude $\sim 75^\circ$ to $90^\circ$ greater or lower than the longitude of the observer. It can thus be complicated to disentangle between ``no emission" and ``non-visible emission", because the observer is not in the beam of the emission. For this, it can be interesting to have multi-point observations, e.g., \correction{including} ground-based \correctiontwo{radio telescopes} such as the Nan\c cay Decameter Array (NDA). \correction{Figures \ref{fig:NDA_observation} displays} observations taken by Juno (\correctiontwo{\ref{fig:NDA_observation}a}) and NDA (\correctiontwo{\ref{fig:NDA_observation}b}) on 2016-12-19. \correction{The observation geometry is shown in Figure \ref{fig:NDA_observation}\correctiontwo{c}, with Juno located at a mean local time of $5.2$~hours, and NDA at a mean local time of $ 12.64$~hours, at the moment of the observations of the radio emissions. Finally, Figure \ref{fig:NDA_observation}\correctiontwo{d} shows the shape of the radio emission as a function of the position of the sources relative to the observer.}

\correction{Multiple \correctiontwo{``B" arcs} are observed by Juno up to almost $30$~MHz, between 11:00 and 12:30} \correctiontwo{\cite<Figure \ref{fig:NDA_observation}a, see also>[who statistically reported these arcs]{2017GeoRL..44.4584I}}. \correctiontwo{The type of the arcs and the position of Juno indicates that the emissions come from the midnight-to-dusk side as seen from Juno (see Figures \ref{fig:NDA_observation}c,d)}.
\correctiontwo{On the other hand, between 09:00 and 09:30, ``A" emissions} are observed by the NDA (Figure \ref{fig:NDA_observation}\correctiontwo{b}) up to almost 30~MHz. \correctiontwo{The type of the emissions seen by the NDA, and its position relative to Jupiter, indicates the emissions come from the dusk side as seen from Earth (see Figures \ref{fig:NDA_observation}c,d)}.
By studying the time delay (e.g., at $24.5$~MHz) between the first emission seen on 2016-12-19T09:08 by the NDA (Earth Time, i.e. $\sim2016$-$12$-$19$T$08$:$23$ Juno Time, taking into account the light travel time) and the first emission seen on 2016-12-19T12:27 (Juno Time) by Juno, \correction{we obtain a $\delta t=4.1$~h}. According to the local time positions of the two observers, \correction{this is consistent with an emission originating from the same source, seen from both side of the beaming cone, and rotating with a sub-corotation rate of $70 \pm 5$~\%, meaning that the source is rotating at $70$~\% of Jupiter's rotation angular frequency} \cite<\correction{taking into account that the emission at $24.5$~MHz is beamed along a hollow cone with aperture angle of $75^\circ \pm 5^\circ$},>[]{2017GeoRL..44.9225L}.

\correction{The beaming angle allowed by the CMI is in the range $75^\circ$--$90^\circ$, and Juno does not see a \correctiontwo{``B"} radio emission before the NDA. Therefore the onset region must be located in a region greater than Juno’s local time plus $75^\circ$--$90^\circ$, and lower than NDA’s local time minus $75^\circ$--$90^\circ$, therefore in the local time range [1110--1740]~$\pm$~0100 hours.}

The lack of emission observed by Juno is therefore partly due to geometry effects, but probably also to a delay in the activation of the sources and in a specific region (dusk). Indeed, the NDA sees an emission before Juno, but no emission is seen by Juno at the previous rotation, indicating that a time delay exists between the compression of the magnetosphere and the activation of newly activated DAM sources. This exact time delay is difficult to determine here, and would require a more statistical study or more observers, but it seems that at least two Jovian rotations are needed before new DAM sources are activated.

\subsubsection{Narrowband kilometric (nKOM) emission}
Finally, the delay for new nKOM emissions to be visible is far \correctiontwo{longer} than for bKOM and DAM emissions. The first new emission appears at $\sim2016$-$12$-$20$T$00$:$00$, i.e. $39$~hours after the first visible bKOM emission. The interval between the peaks in the integrated intensity is not regular, and varies between $\sim9$~h~$14$~min, $\sim10$~h~$22$~min and $\sim9$~h~$44$~min. A closer look to the intensity peaks at different frequencies (see Figure S5b) shows that \correction{the} signal at lower frequencies (e.g., from $70.862$~kHz to $112.43$~kHz) is triggered before the signal at higher frequencies (e.g., at 126.16~kHz and 141.54~kHz), and then disappears first. The interval between the peaks seems to be different depending on the frequency\correction{, which implies different source locations (see \correctiontwo{Section \ref{subsec:radio_response_to_2nd_comp}}, and Section \ref{sec:discussion}, for more details)}.


\subsection{Response of the auroral radio emission to the second compression}
\label{subsec:radio_response_to_2nd_comp}

As mentioned at the beginning of this section, \correction{the dynamic pressure of the solar wind during the second compression event is potentially weaker than during the first event. This is suggested by both (i) the position of the magnetopause, further away from Jupiter (see the second dotted black line at $2016$-$12$-$21$T$08$:$48$), and (ii) the time spent in the magnetosphere which is shorter than during the first event.}

The inspection of the radio emission time series shows that one DAM emission is observed at $\sim2016$-$12$-$21$T$21$:$30$, also observed one rotation later with greater intensity. This emission is \correction{most likely} the reactivation of previously observed sources (as observed during the first compression event). Indeed DAM emission with decreasing intensity is observed $\sim20$~hours before ($\sim2016$-$12$-$21$T$01$:$30$) with the same shape. \correction{Since the} NDA \correction{is} observing only one third of the time we have no contemporaneous observations for this event.

New bKOM emission sources are activated at $\sim2016$-$12$-$21$T$08$:$00$. However, in contrast to the first event, fewer bKOM sources seem to have been activated, since the bKOM emission is not visible at all times, and the sources are activated for a shorter period of time (only visible for $\sim30$~hours vs. $\sim60$~hours).

Finally, regarding the nKOM emission, new nKOM emissions are activated, starting at $\sim2016$-$12$-$22$T$15$:$00$, and lasting for $\sim40$~hours (same duration as for the first compression), with integrated intensity higher than for the first event. This time, the delay between the activation of the bKOM and the nKOM emissions is only $\sim31$~hours.
Again, it can be seen that the period between the peaks in the integrated intensity is not regular. It varies between $\sim10$~h~$30$~min, $\sim9$~h~$50$~min and $\sim10$~h~$54$~min. A closer look to the intensity peaks at different frequencies (see Figure S5c) shows that the signal is first triggered at the lowest frequencies before being triggered at the highest frequencies. Then the signal disappears, or fades, in the same order. The interval between two peaks is different depending on the frequency. Focusing on distribution peaks at each frequency, it can be seen that periodicity increases with decreasing \correctiontwo{frequency. When} the new nKOM emissions are activated, all peaks are almost centered at the same time ($\sim2016$-$12$-$22$T$15$:$45$); one rotation later, the peaks are distributed in order of decreasing frequency, with the 141.54~kHz signal seen first and the 89.172~kHz signal peak seen last. This could be explained by the fact that the lower frequency nKOM is generated at lower density, hence, larger radial distances from Jupiter: the deviation from rigid co-rotation would be greater farther from the planet, \correction{and} the periodicity should be \correctiontwo{longer}.

\section{Summary, Discussion and Perspectives}
\label{sec:perspectives}
\label{sec:summary}
\label{sec:discussion}

\begin{table}
\centering
\caption{Table summarising the results of the study of the response time of radio emissions to compression, as seen by Juno.
For each compression, the dynamic pressure of the solar wind \cite<determined from the model of >[]{2002JGRA..107.1309J}, the type of shock \correction{\cite<determined from the model of >[]{2005JGRA..11011208T}}, the response time of each component of the radio emission (main band of the bKOM, low frequency extension (LFE) of the bKOM, DAM and nKOM) and the activation time (as seen by Juno) are given.}
\resizebox{1.0\textwidth}{!}{\begin{tabular}{|c|c|c|cc|c|c|}
\hline
\hline
\multirow{2}{*}{Compression}                         & Pdyn                       & Type of shock & \multicolumn{2}{|c|}{\multirow{2}{*}{Auroral radio emission}} & \multirow{2}{*}{Activation time} & \multirow{2}{*}{Duration} \\
                                                     & \citeA{2002JGRA..107.1309J}      & \citeA{2005JGRA..11011208T}  & \multicolumn{2}{|c|}{}                      &          &            \\
\hline
{\multirow{4}{*}{1st compression}} & \multirow{3}{*}{0.70}  & \multirow{3}{*}{FRS}              & \multirow{2}{*}{bKOM}                   &        Main band           &         $\le 10$s~min            &   $\sim60$~hours \\
\multicolumn{1}{|c|}{}                                 &                       &                                &                                         &          LFE         &        $\sim34$ hours             &       $1$~h~$15$~min                \\
\multicolumn{1}{|c|}{}                                 &                       &                                & \multicolumn{2}{|c|}{DAM}                                                  &          $\sim28$~hours          &     $\sim30$~hours                \\
\multicolumn{1}{|c|}{}                                 &                       &                                & \multicolumn{2}{|c|}{nKOM}                                                  &          $\sim39$~hours          &     $\sim$40~hours                \\
\hline
\multirow{4}{*}{2nd compression}                     & \multirow{3}{*}{0.48} & \multirow{3}{*}{FFS}              & \multirow{2}{*}{bKOM}                   &         Main band         &       $\le 10$~min                &            $\sim30$~hours         \\

                                                     &                       &                        &                                         &       LFE            &    $\le 10$~min                  &       $\sim15$~hours              \\
                                                     &                       &                                 & \multicolumn{2}{|c|}{DAM}                   &        $\sim12$~h~$45$~min             &  $10$~hours                     \\
\multicolumn{1}{|c|}{}                                 &                       &                                & \multicolumn{2}{|c|}{nKOM}                                                  &          $\sim31$~hours          &     $\sim40$~hours                \\
\hline
\hline
\end{tabular}}
\label{table:table_resume}
\end{table}

In this paper, we have presented in Section \ref{sec:methodo} a \correction{set} of magnetospheric boundary crossings (See Figure \ref{fig:MS_boundary_crossing}). \correction{More detailed information on each crossing, such as their exact time, their positions in different coordinate systems, and several added values ($P_\mathrm{dyn}$, magnetopause and bow shock standoff distances) are given in Supporting Information (Tables S1, S2), as well as statistical distributions for these added values (Figure S2). The files corresponding to Tables S1, S2} are accessible \correctiontwo{through \citeA{2022_boundary_crossings_lists}}.

\correction{In Section \ref{sec:results}, we presented} case studies of the response of Jovian radio emission to \correction{strong to moderate} magnetospheric compressions, inferred by magnetopause crossings. Using \correctiontwo{the} \citeA{2002JGRA..107.1309J} model, we calculated the dynamic pressure (lower limit) of the solar wind (see Table \ref{table:table_resume}), and its main characteristics and type of shocks associated with these events using the \citeA{2005JGRA..11011208T}. We determined that the first magnetopause crossing is potentially due to (i) either a stronger and shorter compression, (ii) or higher solar wind dynamic pressure, based on the time spent by Juno in the magnetosheath.

We chose to study the magnetopause crossings occurring between 2016-12-17T00:00 and 2016-12-24T04:15 (fourth orbit of Juno). These magnetopause crossings are among the innermost cases (see Figure 1a and S1a), corresponding to strong compressions ($P_\mathrm{dyn} \subset$ [$0.5$-$0.7$] according to the \citeA{2002JGRA..107.1309J} model). These compressions occur when Juno is still on the dawn side of the magnetosphere, i.e. in a region where the model of \citeA{2002JGRA..107.1309J} is valid, in contrast to the dusk side where it is less constrained.  Moreover, during this 7-day interval, we observe several magnetopause crossings, which can be grouped into 2 phases of magnetospheric compression. These two cases also seem to correspond to two different types of shock: FFS and FRS, according to the propagation model of \citeA{2005JGRA..11011208T}, with different responses observed in the radio components (see Table \ref{table:table_resume}). 

Concerning the \correction{radio emission response to the compressions}, we have determined that the bKOM sources \correctiontwo{are} the first to be triggered, at almost every longitude, almost immediately after the observation of the first magnetic disturbances and density perturbations. The bKOM emission is then observed \correction{over} $60$~hours for the first compression and \correction{for} $30$~hours for the second one. Low Frequency Extension\correction{s}, i.e. emissions going down to $20$~kHz, are observed in both case\correction{s} for a shorter duration.

In both case\correction{s}, the DAM emissions are the second ones to be observed, at least one rotation after the start of the compression, and only in the noon-dusk sector, \correction{i.e. inside the local time range [1110--1740]}. \correction{This sector includes that determined by \citeA{2012P&SS...70..114H, 2014P&SS...99..136H}, but is necessarily less precise given that we are only studying two cases here. A statistical study with Juno will provide further constraints, given the evolution of Juno's local time position during its mission.} Our results seem to show that both FRS and FFS activate new, or re-activate, DAM emissions on \correction{the} dusk side only. \correction{This} is partially in agreement with \citeA{2012P&SS...70..114H, 2014P&SS...99..136H}, who showed that FFS mainly trigger DAM emission on the dusk side, while FRS trigger emissions on the dusk and dawn sides. However, since we are \correction{measuring} radio emission only above $3.5$~MHz in this study (due to Waves sensitivity) we are missing part of the DAM and most of the HOM emissions, that can go down to $0.3$~MHz, while \citeA{2012P&SS...70..114H, 2014P&SS...99..136H} used Cassini radio measurements, down to $0.1$~MHz. The DAM emission lasts for $30$~hours in the first case, \correction{and} $10$~hours in the second case. In both cases, sources rotate in subcorotation, with a \correction{rate of $70 \pm 5$~\% of rigid corotation. This value is comparable with the values obtained by \citeA{2012P&SS...70..114H, 2014P&SS...99..136H}.}

Concerning the activated nKOM emissions, we observe a strong difference compared to the bKOM and DAM emissions, with a long delay between compression and activation of the nKOM sources ($\sim30$ to $40$~hours). nKOM emission is then observed for $\sim40$~hours in both compression events. The periodicity of the nKOM peaks is frequency-dependent and increases with decreasing frequency. This would be related to the mechanism, producing emissions at the plasma frequency which is proportional to the local plasma density. \correction{Therefore,} low-frequency emissions \correction{are} produced f\correction{a}rther from Jupiter than higher-frequency emissions. The activation of new nKOM sources seems related to the relaxation/reconfiguration phase of the magnetosphere. As these emissions are produced by different mechanisms, it is not surprising that the activation of these emissions is also different. However, it is possible that the energetic events observed by \citeA{1998GeoRL..25.2905L, 2016JGRA..121.9651L} could be caused or amplified by \correction{and expansion} of the magnetosphere, which would amplify the centrifugal ejection of matter. It will therefore be mandatory to study in detail the nKOM during plasmasheet distortion, which will require a list of magnetic disturbances measured during plasma sheet \correction{crossings}, simultaneously to compression events. But this is beyond the scope of this current article, and will be the subject of an upcoming study.

To get a better estimate of the conditions in the solar wind, such as the solar wind dynamic pressure and velocity, the \citeA{2019JGRA..124.7799T} analytical method could be used, based on Juno/JADE measurements inside the magnetosheath (Juno/JADE data were not available for the event studied in Section \ref{sec:results}). This will be compared to estimation of the dynamic pressure obtained from \citeA{2002JGRA..107.1309J} magnetosphere boundaries model and \citeA{2005JGRA..11011208T} propagation tool model. We could also use different solar wind propagation tools, \correctiontwo{such as ``HuXT" model \cite<Heliospheric Upwind Extrapolation with time dependence>[]{2020SoPh..295...43O}}, ``WSA-ENLIL solar wind simulation'', \correctiontwo{``HelioCast'' \cite{2023JSWSC..13...11R}} or the ``CDPP/Propagation Tool'' extended to Jupiter \cite{2017P&SS..147...61R}.

To go further on the generalization of the response of Jovian radio emissions, the activation of new sources or the amplification of existing radio emissions, and their intensity to magnetospheric compression and solar wind characteristics  (dynamic pressure, velocity, temperature, magnetic field orientation), a statistical study will be necessary. The same method will be used and will be applied to all the compression events determined from the list of magnetopause crossings provided in the SI tables (see also \correctiontwo{Figures} \ref{fig:MS_boundary_crossing} and SI1). \correctiontwo{This will involve using boundary crossings to infer compressions, examining the response of associated radio emissions, and grouping case studies by properties such as solar wind dynamic pressure, or shock type.}

\correction{There are several benefits to a future statistical study. The first is to explore the differences between dawn and dusk side responses, and the different properties of the boundaries of the magnetopshere \cite<e.g., Kelvin-Helmholtz instability,>[]{2021GeoRL..4894002M}, or the differences in the observation of radio sources (beaming constraints).
The second aspect is the opportunity to explore different classes of behaviour in terms of magnetosperic compression state. Due to the precession of the apojoves, we observe the compression of the magnetosphere from different positions in the magnetosphere. As shown in Figure \ref{fig:MS_boundary_crossing}c, the nature of the boundary motion is highly variable, and the number of boundary crossings varies greatly from one orbit to another. Some orbits have clean boundary crossings, while other orbits have multiple crossings in a short time.  This makes it possible to study the radio response during the compression and relaxation phases, but also during the stationary state - see Figures \ref{fig:crossing_examples}c,f for an example.
Thirdly, the long period of time between Juno's insertion into Jovian orbit (July 2016) and the \correctiontwo{latest} orbits of the extended mission \correctiontwo{(perijoves~ $\geq 50$)} covers \correctiontwo{two different phases of two different} solar cycle and \correctiontwo{different} Jovian seasons, which \correctiontwo{could} allow us to explore the response of radio emissions to compression as a function of the solar cycle\correctiontwo{s} and Jovian seasons.}

\correction{At the time of writing, Juno is still crossing the boundaries on the high southern latitude dusk side, and thus a full statistical exploration of the broad parameter space should await the completion of these apojove passes. Moreover, the comprehensive labelled radio emissions catalogue \cite{2021_Juno_Waves_catalog} is currently being updated to cover the whole mission.}


\acknowledgments
C.~K.~Louis', C.~M.~Jackman's, A.~R.~Fogg's and S.~C.~McEntee's work at the Dublin Institute for Advanced Studies was funded by the Science Foundation Ireland Grant 18/FRL/6199. The research at the University of Iowa is supported by NASA through Contract  699041X with Southwest Research Institute. D.~M.~Weigt’s work at the Dublin Institute for Advanced Studies was funded by European Union’s Horizon 2020 research and innovation programme under Grant agreement No. 952439 and project number AO 2-1927/22/NL/GLC/ov as part of the ESA OSIP Nanosats for Spaceweather Campaign D.~M.~Weigt's work at Aalto University was funded from the European Research Council (ERC) under the European Union’s Horizon 2020 research and innovation programme (project ``SYCOS", grant agreement n$^o$ 101101005). The research at the University of Iowa is supported by NASA through Contract 699041X  with the Southwest Research Institute. WSK acknowledges the use of the Space Physics Data Repository at the University of Iowa supported by the Roy J. Carver Charitable Trust.

\section*{Data Availability Statements}
The Juno/Waves dataset displayed in this paper, produced by \citeA{2021JGRA..12629435L}, is accessible at \url{https://doi.org/10.25935/6jg4-mk86} \cite{2021Juno_Waves_Calibrated_data_collection}, and the catalogue can be download at \url{https://doi.org/10.25935/nhb2-wy29} \cite{2021_Juno_Waves_catalog} .
The Juno/MAG magnetic field data are accessible through the NASA/PDS website \cite{2017_FGM_pds_Connerney}. Figure \ref{fig:MS_boundary_crossing} was produced using the Jupiter magnetosphere boundaries crossings given in the SI Tables S1 and S2 \cite{2022_boundary_crossings_lists}. Juno/Waves and Juno/\correctiontwo{MAG} data were displayed using the Autoplot tool \cite{2010sdfghF}. The Nan\c cay Decameter Array dataset displayed in Figure \ref{fig:NDA_observation} is accessible at \url{https://doi.org/10.25935/PBPE-BF82} \cite{2021_NDA_data_collection_Lamy}. The routine that allows to determine the dynamic pressure from the \citeA{2002JGRA..107.1309J} model are accessible at \url{https://github.com/DIASPlanetary/jupiter_magnetosphere_boundaries}. Juno ephemeris and MAG data (in JSO coordinates system) were retrieved from \url{http://amda.cdpp.eu/} \cite{2021P&SS..20105214G}. Juno ephemeris used to inferred the dynamic pressure (in JSS coordinate) were retrieved from \url{https://wgc.jpl.nasa.gov:8443/webgeocalc/#StateVector}.


%
%

\bibliography{Louis_2022_radio_SW.bib}

%
%
%
%
%

\end{document}